\documentclass[pra,a4paper,twocolumn,showpacs,superscriptaddress]{revtex4}
\usepackage{amsmath}
\usepackage{amsfonts}
\usepackage{bbm}
\usepackage{graphicx}
\usepackage{longtable}
\begin{document}

\title{Quantum which-way information and fringe visibility when the detector is entangled with an ancilla} 
\author{J. Prabhu Tej}
\affiliation{Department of Physics, Bangalore University, 
Bangalore-560 056, India}
\author{A. R. Usha Devi}\email{arutth@rediffmail.com} 
\affiliation{Department of Physics, Bangalore University, 
Bangalore-560 056, India}
\affiliation{Inspire Institute Inc., Alexandria, Virginia, 22303, USA.}
\author{H. S. Karthik}
\affiliation{Raman Research Institute, Bangalore 560 080, India} 
\author{Sudha} 
\affiliation{Department of Physics, Kuvempu University, Shankaraghatta, Shimoga-577 451, India.}
\affiliation{Inspire Institute Inc., Alexandria, Virginia, 22303, USA,}
\author{A. K. Rajagopal}
\affiliation{Inspire Institute Inc., Alexandria, Virginia, 22303, USA.}
\affiliation{Harish-Chandra Research Institute, Chhatnag Road, Jhunsi, Allahabad 211 019, India.}
\affiliation{Institute of Mathematical Sciences, C.I.T Campus, Tharamani, Chennai 600 113, India  }
\date{\today}
\begin{abstract} 
Quantum mechanical wave-particle duality is quantified in terms of a trade-off relation between the fringe visibility and the which-way distinguishability in an interference experiment. This relation is recently generalized by Banaszek et. al., (Nature Communications {\bf 4}, 2594, (2013)) when the particle is equipped with an internal degree of freedom such as spin. Here, we extend the visibility-distinguishability trade-off relation to quantum interference of a particle possessing  an internal degree of freedom,  when the {\em which-way} detector state is entangled with an ancillary system.  We introduce an {\em extended which-way distinguishability} ${\cal D}_E$ and the associated {\em extended fringe visibility} ${\cal V}_E$  satisfying the inequality ${\cal D}^2_E+{\cal V}^2_E\leq 1$ in this scenario. We illustrate, with the help of three specific examples, that while the which-way information inferred solely from the detector state (without ancilla) vanishes, the {\em extended distinguishability} retrievable via  measurements  on the  detector-ancilla entangled state is non-zero. Furthermore, in all the three examples, the {\em extended visibility} and the {\em generalized visibility} (which was introduced by Banaszek et. al., Nature Communications {\bf 4}, 2594, (2013)) match identically with each other.    
\end{abstract}
\maketitle

\section{Introduction}

Visibility of fringes in a single quantum particle interference experiment sets limits on the which-way information~\cite{WKW, Sc, GY, En}, thus demonstrating wave-particle duality. Very recently, Banaszek et. al.~\cite{Ban} analyzed the trade-off between interference visibility and which-path distinguishability for a quantum particle possessing an internal structure (such as spin or polarization). In this setting, an interaction of the  internal spin state with the detector system is shown to offer non-trivial identifications. The internal structure could play a manipulative role in controlling the information about which-path in the interferometer arms is taken by the particle. Trade-off between the amount of which-way information encoded in the detector system and the fringe visibility is  captured in terms of a generalized complementarity relation~\cite{Ban}, by extending the notion of fringe visibility in terms of the internal spin states as well as their interaction with the detector.  

To place things in order, we outline the basic scenario,~\cite{En} where a single quantum particle (quanton) $Q$ travelling through a two-path interferometer (double slit or Mach-Zehnder interferrometer), with the paths being equiprobable. Let us denote the initial state of the quanton and the detector system by $\rho^{\rm (in)}_{QD}=\rho^{\rm (in)}_Q\otimes \rho^{\rm (in)}_D$. When  the quanton takes either path $0$ or $1$ of the interferometer arms,  the detector state  correspondingly gets transformed into    
\begin{eqnarray}
\rho^{(i)}_D&=U^{(i)}_D \rho^{\rm (in)}_{D} U^{(i)\dag}_D,\ \ i=0,1,
\end{eqnarray}
where $U^{(i)}_D$  denote unitary  transformations on the detector states corresponding to the paths of the quanton. (The interaction
is constrained such that the quanton paths  cannot get transferred into one another due to interaction. The  final detector state  after the interaction is then  given by,  $\rho^{\rm (fin)}_D=\frac{1}{2}\, \rho^{(0)}_D+ \frac{1}{2}\,\rho^{(1)}_D$). 
 
Which-way information is quantified in terms of {\em distinguishability} $0\leq {\cal D}\leq 1$, which is the trace-distance between the detector states 
$\rho^{(0)}_D$ and $\rho^{(1)}_D$: 
\begin{eqnarray}
{\cal D}&=&\frac{1}{2}\, \vert\vert \rho^{(0)}_D-\rho^{(1)}_D\vert\vert. 
\end{eqnarray}  
(Here, $\vert\vert\, A\, \vert\vert={\rm Tr}\,[\sqrt{A^\dag\,A}]$ denotes the trace-norm of $A$).   

It may be noted that the distinguishability is the maximum of the difference of probabilities of the correct and incorrect decisions about the paths~\cite{ Helstrom}.  The paths of the quanton cannot be distinguished when ${\cal D}=0$ (i.e.,when  $\rho^{(0)}_D\equiv \rho^{(1)}_D$) whereas, they are perfectly distinguishable when ${\cal D}=1$ (i.e., when $\rho^{(0)}_D$ and $\rho^{(1)}_D$ are orthogonal). 
Consequently, the fringe visibility ${\cal V}$ is given by~\cite{En} 
\begin{equation}
{\cal V}=\left\vert {\rm Tr}[U^{(0)}_D\, \rho^{\rm (in)}_{D}\, U^{(1)\dag}_D] \right\vert.   
\end{equation}
Visibility $0\leq {\cal V}\leq 1$ characterizes the ability of the quanton, distributed between two paths, to interfere after geting combined at the exit of the interferometer. Wave-particle duality in the quantum interference experiment is expressed in terms of the trade-off relation~\cite{En} between visibililty and distinguishability:  
\begin{equation} 
{\cal D}^2+{\cal V}^2\leq 1. 
\end{equation}

In a more general set up, explored recently~\cite{Ban}, the quanton is equipped with an internal degree of freedom such as spin (characterized by a $d_S$ level quantum system) and  it is recognized that there is an intricate relation between  the which-way information ${\cal D}$ on the initial preparation of the internal spin state, in addition to  the specific details  of its interaction with the detector. Banaszek et. al.~\cite{Ban} demonstrated a stringent bound on distinguishability  in terms of {\em generalized fringe visibility}, which depends on the initial preparation of spin state as well as on the nature of its interaction with the detector system. The generalized trade-off inequality reads as~\cite{Ban}, 
\begin{equation}
\label{dvg}
{\cal D}^2+{\cal V}_G^2\leq 1 
\end{equation}  
where the distinguishability  ${\cal D}=\frac{1}{2}\, \vert\vert \rho^{(0)}_D- \rho^{(1)}_D\vert\vert$ captures the leak-out of which-way information to the detector (here,   $\rho^{(i)}_D=2\, _Q\langle i\vert {\rm Tr}_S[U_{QSD}\, \rho^{\rm (in)}_{QS}\otimes \rho^{\rm (in)}_{D}\, U^\dag_{QSD}]\vert i\rangle_Q,\ \ i=0,1$ are the detector states corresponding to the quanton paths);   the unitary interaction $U_{QSD}$  is constrained to be of the form $U_{QSD}=\sum_{i=0,1}\, \vert i\rangle_Q\langle i\vert \otimes U^{(i)}_{SD}$ such that the which-way interaction does not shift the quanton between interferometer arms.  

The generalized fringe visiblity ${\cal V}_G$ in (\ref{dvg}) is given by~\cite{Ban},  
\begin{equation}
\label{vg} 
{\cal V}_G= d_S\, \vert\vert (\mathbbm{1}\otimes \Lambda_{01})[(I_S\otimes \sqrt{\rho^{\rm (in)}_{S0}})\vert\Phi_{+}\rangle\langle \Phi_{+}\vert\, (I_S\otimes \sqrt{\rho^{\rm (in)}_{S1}})]\vert\vert,
\end{equation} 
where, the unitary interaction of the detector with the spin subsystem corresponds to the 
action of a quantum channel~\cite{NC} $\Lambda$ on the internal spin state as explained below: Let $\rho^{\rm (in)}_{QSD}=\rho^{\rm (in)}_{QS}\otimes \rho^{\rm (in)}_D$ denote the initial  quanton path-spin  (denoted by $QS$) and the detector (denoted by $D$) states. Unitary interaction between the detector and the internal spin results in the final state $\rho^{\rm (fin)}_{QSD}=U_{QSD}\, \rho^{\rm (in)}_{QSD}\, U^\dag_{QSD}$. This unitary interaction on the initial state $\rho^{\rm (in)}_{QSD}$ may be viewed as the action of a quantum  superoperator  $\Lambda$ on the input state $\rho^{\rm (in)}_{QS}$ i.e., $\Lambda(\rho^{\rm (in)}_{QS})={\rm Tr}_D[\rho^{\rm (fin)}_{QSD}]$. Further, as the quanton does not get switched between the interferometer arms $0$ and $1$ as a result of the interaction, the channel $\Lambda$ must be of the form: 
\begin{equation} 
\Lambda(\vert i\rangle_Q\langle j\vert \otimes \sigma_S)=\vert i\rangle_Q\langle j\vert \otimes \Lambda_{ij}(\sigma_S),\ \ \ \  i,j=0,1
\end{equation}     
where $\sigma_S$ corresponds to any operator in the spin space. Further, in (\ref{vg}), the states $\rho^{\rm (in)}_{S0}, \ \rho^{\rm (in)}_{S1}$ are the initial  spin states along the paths $0$, $1$ and are given by $\rho^{\rm (in)}_{S0}=2\, _Q\langle 0\vert \rho^{\rm (in)}_{QS}\vert 0\rangle_Q, \ 
\rho^{\rm (in)}_{S1}=2\, _Q\langle 1\vert \rho^{\rm (in)}_{QS}\vert 1\rangle_Q$; $I_S$ denotes identity operator in the spin space and $\mathbbm{1}$ denotes the identity channel; the state $\vert \Phi_{+}\rangle=\frac{1}{\sqrt{d_S}}\,\displaystyle\sum_{\alpha=0}^{d_S-1} \vert \alpha\rangle_S  \, \vert \alpha\rangle_{S'}$ is a maximally entangled state of two replicas of the spin system.

Banaszek et.al.,\cite{Ban} analyze  specific examples to demonstrate the intricate role played by the internal spin  state preparation corresponding to  specific interaction channels $\Lambda_{01}$.  Specifically, when there is no interaction with the detector, the channel reduces to  $\Lambda_{01}=\mathbbm{1}$ and after simplification of (\ref{vg}) one obtains  ${\cal V}_G=\sqrt{{\rm Tr}[\rho^{\rm (in)}_{S0}]\, {\rm Tr}[\rho^{\rm (in)}_{S1}]}=1$, irrespective of the preparation of the initial spin state  (i.e., when the detector gains no information about the path, visibility ${\cal V}_G$ is 1  and the distinguishability ${\cal D}$ is 0). When the interaction channel is given by $\Lambda_{01}(\sigma_S)={\rm Tr}[\sigma_S]\, \Sigma_S$, with $\Sigma_S$ being a fixed unit-trace hermitian operator, the generalized visibility reduces to the fidelity~\cite{U, J} between the spin states $\rho^{\rm (in)}_{S0},\ \rho^{\rm (in)}_{S1}$ i.e., ${\cal V}_G={\rm Tr}[\sqrt{\sqrt{\rho^{\rm (in)}_{S0}}\, \rho^{\rm (in)}_{S1}\, \sqrt{\rho^{\rm (in)}_{S0}}}]=F(\rho^{\rm (in)}_{S0}, \rho^{\rm (in)}_{S1})$. Thus,  the which-way information can be blocked by preparing identical spin states for both the paths i.e., $\rho^{\rm (in)}_{S0}=\rho^{\rm (in)}_{S1}$,  so that the generalized visibility takes its maximum value 1. In yet another interesting example of the interaction channel, defined through  $\Lambda_{01}(\sigma_S)=\sigma^T_S/d_S$ (where $\sigma^T_S$ is the transpose of the spin operator $\sigma_S$), the generalized visibility (\ref{vg}) gets simplified to ${\cal V}_G = \vert\vert \sqrt{\rho^{\rm (in)}_{S0}}\vert\vert\, \vert\vert \sqrt{\rho^{\rm (in)}_{S1}}\vert\vert / d_S$.  The spin states in both the paths,  prepared initially in a completely mixed state $\rho^{\rm (in)}_{S0}=
\rho^{\rm (in)}_{S1}=I_S/d_S$, would lead to the generalized fringe visibility ${\cal V}_G=1$ and hence, the  which-path information to the detector can be blocked. 

In the present work, we explore the trade-off between the  fringe visibility and the which-way information retrievable, when the  detector is entangled with an  ancilla.  Basically, the which-way information corresponds, in particular, to the discrimination of the detector states $\rho^{(0)}_D$ and $\rho^{(1)}_D$,    when the quanton chooses path 0 or path 1. One may view the states  $\rho^{(0)}_D$ and $\rho^{(1)}_D$ as the outputs of  completely positive, trace preserving quantum channels~\cite{note} $\Phi_0$, $\Phi_1$,  acting on the input state $\rho^{\rm (in)}_D$.   The problem of gaining which-way information via distinguishing the two detector states $\rho^{(0)}_D$ and $\rho^{(1)}_D$ can then be linked with that of  discriminating the two quantum channels $\Phi_0$, $\Phi_1$. A useful measure of distinguishability of two quantum channels is given by their  trace distance, 
\begin{equation}
\label{chd}
\frac{1}{2}\vert\vert \Phi_0-\Phi_1\vert\vert=\stackrel{\max}{_\rho} \,\frac{1}{2}\, \vert\vert \Phi_0(\rho)-\Phi_1(\rho)\vert\vert
\end{equation}
 where the maximum is taken over all  pure input states $\rho$ ~\cite{War, Sacchi1}. However, an optimal approach to maximize the observable difference  between the two channels is to prepare the input state entangled with another auxiliary system; apply  one of the channels (chosen randomly) to the input state (with ancillary subsystem being an idler) and then measure the resulting output bipartite state to identify which of the channels was applied. It has been established that channel inputs, which are entangled with an ancilla could offer remarkable improvements in distinguishing some pairs of channels~\cite{War, Sacchi1, kit, chi, Dar, Acin, Sacchi2, SL, Tan, ARU, Piani}. In fact, there are examples of channels~\cite{War} that can be  distinguished {\em perfectly}, when they are applied to one part of a  maximally entangled state, while they are {\em indistinguishable} if the auxilliary system is not employed.    

\section{Which-way information using entangled detector-ancilla state}

Our purpose here is to investigate the enhancement of  which-way information,  given that the detector  is entangled with an  ancilla $D'$ of  same dimension as that of the detector system~\cite{note2}. We define {\em extended distinguishability} achievable with an entangled detector-ancilla initial state  by, 
\begin{eqnarray}
{\cal D}_E&=&\frac{1}{2} \vert\vert (\Phi_0\otimes \mathbbm{1})(\rho^{\rm (in)}_{DD'})- (\Phi_1\otimes \mathbbm{1})((\rho^{\rm (in)}_{DD'})\vert\vert \nonumber \\
\label{dg}
 &=& \frac{1}{2}\, \vert\vert \rho^{(0)}_{DD'}-\rho^{(1)}_{DD'}\vert\vert
\end{eqnarray}  
where 
\begin{eqnarray*}
&&\rho^{(i)}_{DD'}= (\Phi_i\otimes \mathbbm{1})(\rho^{\rm (in)}_{DD'})\nonumber \\ 
&&\ \ \ \ = 2\, _Q\langle i\vert {\rm Tr}_S\,[U_{QSD}\otimes I_{D'}\, \rho_{QS}\otimes \rho_{DD'}\, U^\dag_{QSD}\otimes I_{D'}]\vert i\rangle_Q, 
\end{eqnarray*}
denote the final detector-ancilla states corresponding to quanton paths $i=0,1$.

Using the inequality~\cite{Fuchs}  $D(\varrho, \tau)\leq \sqrt{1-F^2(\varrho,\, \tau)},$ relating the trace distance $D(\varrho,\, \tau)=\frac{1}{2}\, \vert\vert \varrho-\tau\vert\vert$ and the fidelity 
$F(\varrho,\, \tau)={\rm Tr}[\sqrt{\sqrt{\varrho}\,\, \tau\,\sqrt{\varrho}}]$ between  the two density operators $\varrho$ and $\tau$, we obtain the following relation for the extended distinguishability:
\begin{equation}
\label{df}
{\cal D}_E\leq \sqrt{1-F^2(\rho^{(0)}_{DD'},\rho^{(1)}_{DD'})}. 
\end{equation} 
Expressing the extended distinguishability as ${\cal D}_E\leq \sqrt{1-{\cal V}^2_E}$, we define the corresponding {\em extended fringe visibility} by  
\begin{eqnarray}
\label{fv}
{\cal V}_E&=&F(\rho^{(0)}_{DD'},\rho^{(1)}_{DD'}). 
\end{eqnarray}

We now proceed to analyze three specific  examples of interaction to illurstrate that  the extended which-way distinguishability ${\cal D}_E$ can assume non-zero values,  even when the  distinguishability  ${\cal D}$ inferred by measuring only the detector states vanishes. In the meanwhile,  we also find that the extended visibility and the generalized visibiilty~\cite{Ban} agree identically with each other in these examples. 

\section{Examples} 

 Let us consider a quanton -- with two dimensional internal spin states  $\vert 0\rangle_S, \vert 1\rangle_S$  -- traveling through a Mach-Zehnder interferometer. We consider a pure entangled  detector-ancilla input state 
\begin{equation}
\label{ee'}
 \vert \Psi\rangle_{DD'}= \frac{1}{\sqrt{2}}\, (\vert 0\rangle_D\, \vert 0\rangle_{D'}+\vert 1\rangle_D\,\vert\, 1\rangle_{D'}). 
\end{equation} 
Here, $\{\vert k\rangle_D,\ k=0,1\}$ and  $\{\vert l\rangle_{D'}, l=0,1\}$) denote  orthogonal states of the detector $D$ and ancilla $D'$). Let 
$\vert 0\rangle_Q$, $\vert 1\rangle_Q$ denote the state of the quanton in path $0$ and  $1$ respectively.  Internal spin states  in paths $0$, $1$ are denoted by 
$\vert \psi_0\rangle_S=a_{00}\, \vert 0\rangle_S+ a_{01}\, \vert 1\rangle_S$  and 
$\vert \psi_1\rangle_S=a_{10}\, \vert 0\rangle_S+ a_{11}\, \vert 1\rangle_S$  (the coefficients $a_{\alpha\, \alpha'}$ obey the normalization condition 
$\displaystyle\sum_{\alpha '=0,1}\, \vert a_{\alpha\, \alpha'}\vert^2=1, \ \alpha=0,1$).  

We consider the unitary interaction on the quanton spin with the detector $D$ along the paths $0$, $1$ to be of the form,  
\begin{eqnarray}
U^{(0)}_{SD}= \left( \begin{array}{llll}
1 & 0 & 0 & 0 \\ 
0 & 1  & 0 & 0 \\
0 & 0  & 0 & 1 \\
0 & 0 & 1 & 0 \end{array}
\right)\ \ 
U^{(1)}_{SD}= \left( \begin{array}{llll}
0 & 1 & 0 & 0 \\ 
1 & 0  & 0 & 0 \\
0 & 0  & 1 & 0 \\
0 & 0 & 0 & 1 \end{array}
\right)
\end{eqnarray} 
when expressed in the basis $\{\vert 0\rangle_S\, \vert 0\rangle_D,\, \vert 0\rangle_S\, \vert 1\rangle_D,$  $\vert 1\rangle_S\, \vert 0\rangle_D,$ 
 $\vert 1\rangle_S\, \vert 1\rangle_D\})$.
Under this unitary interaction with the detector  the initial  quanton path-spin state  $\vert \zeta\rangle_{QS}=\frac{1}{\sqrt{2}}\, (\vert 0\rangle_Q\, \vert \psi_0\rangle_S+\vert 1\rangle_Q\, \vert \psi_1\rangle_S)$ and the detector-ancilla state $\vert \Psi\rangle_{DD'}$ (see (\ref{ee'})) get transformed to $\vert \zeta\rangle_{QS}\, \vert \Psi\rangle_{DD'}  \rightarrow \vert\varphi\rangle_{QSDD'}$, which is given explicitly by, 
\begin{eqnarray}
&&\vert \varphi\rangle_{QSDD'}=\frac{1}{\sqrt{2}}\, \left[(a_{00} \vert 0\rangle_Q\, \vert 0\rangle_S\, + a_{11}\, \vert 1\rangle_Q\, \vert 1\rangle_S)\,
\vert \Psi\rangle_{DD'} \right. \nonumber \\ 
&&\left. \ \ \ \ \   + (a_{01}\, \vert 0\rangle_Q\, \vert 1\rangle_S\, +a_{10} \vert 1\rangle_Q\, \vert 0\rangle_S) \, \vert \Psi_\perp\rangle_{DD'} \right] 
\end{eqnarray}
where $\vert\Psi_\perp\rangle_{DD'}=\frac{1}{\sqrt{2}}\, (\vert 0\rangle_D\, \vert 1\rangle_{D'}+\vert 1\rangle_D\,\vert\, 0\rangle_{D'})$. The quanton path-spin final density operator $\rho^{\rm (fin)}_{QS}~=~{\rm Tr}_{DD'} [\vert \varphi\rangle_{QSDD'}\langle \varphi\vert]$ is then found to be, 
\begin{eqnarray}
\label{finqs}
&&\rho^{\rm (fin)}_{QS} = \frac{1}{2}\left[\vert 0\rangle_Q\langle 0\vert \otimes ( \vert a_{00}\vert^2\, \vert 0\rangle_S\langle 0\vert+ 
\vert a_{01}\vert^2\, \vert 1\rangle_S\langle 1\vert) \right.\nonumber \\ 
&&\ \  \ \ +\vert 1\rangle_Q\langle 1\vert \otimes (\vert a_{10}\vert^2\, \vert 0\rangle_S\langle 0\vert+ \vert a_{11}\vert^2\, \vert 1\rangle_S\langle 1\vert)   \nonumber \\ 
 &&  \ \ \ \  + \vert 0\rangle_Q\langle 1\vert \otimes ( a_{00}\, a^{*}_{11}\, \vert 0\rangle_S\langle 1\vert+ a_{01}\, a^*_{10}\, \, \, \vert 1\rangle_S\langle 0\vert) \nonumber \\
&&\left. \ \ \ \  +\vert 1\rangle_Q\langle 0\vert \otimes ( a^*_{00}\, a_{11}\, \vert 1\rangle_S\langle 0\vert+ a^*_{01}\,a_{10}\,  \, \vert 0\rangle_S\langle 1\vert)  \right]. 
\end{eqnarray} 

The states  $\rho^{(i)}_{DD'},\, \ i=0,1 $  of the detector-ancilla system  after the interaction are obtained by 
$\rho^{(i)}_{DD'}=\   _Q\langle i\vert {\rm Tr}_{S}[\vert \varphi\rangle_{QSDD'}\langle\varphi\vert]\, \vert i\rangle_Q,$ and  are explicitly  given (in the basis 
$\{\vert 0\rangle_D\, \vert 0\rangle_{D'},\, \vert 0\rangle_D\, \vert 1\rangle_{D'}, \vert 1\rangle_D\, \vert 0\rangle_{D'}, 
\vert 1\rangle_D\, \vert 1\rangle_{D'}\})$  by 
\begin{eqnarray*}
\rho^{(0)}_{DD'}&=& \frac{1}{2}\,  \left( \begin{array}{llll}
\vert a_{00}\vert^2 & 0 & 0 & \vert a_{00}\vert^2\\ 
0 & \vert a_{01}\vert^2  & \vert a_{01}\vert^2 & 0 \\
0 & \vert a_{01}\vert^2  & \vert a_{01}\vert^2 & 0 \\
\vert a_{00}\vert^2 & 0 & 0 & \vert a_{00}\vert^2 \end{array}, 
\right), \\ 
\rho^{(1)}_{DD'}&=& \frac{1}{2}\,  \left( \begin{array}{llll}
\vert a_{11}\vert^2 & 0 & 0 & \vert a_{11}\vert^2\\ 
0 & \vert a_{10}\vert^2  & \vert a_{10}\vert^2 & 0 \\
0 & \vert a_{10}\vert^2  & \vert a_{10}\vert^2 & 0 \\
\vert a_{11}\vert^2 & 0 & 0 & \vert a_{11}\vert^2 \end{array}
\right).
\end{eqnarray*}

The which-way information retrievable from the {\em extended distinguishability} (see (\ref{dg})) is given by, 
\begin{equation}
\label{1de}
{\cal D}_E=\left\{ \begin{array}{l} (\vert a_{00}\vert^2-\vert a_{11}\vert^2)\ \ {\rm if}\ \ \vert a_{00}\vert^2> \vert a_{11}\vert^2 \\
(\vert a_{11}\vert^2-\vert a_{00}\vert^2)\ \ {\rm if}\ \ \vert a_{11}\vert^2> \vert a_{00}\vert^2 \end{array}
   \right. 
\end{equation} 
and ${\cal D}_E=0$ when $\vert a_{00}\vert^2 =\vert a_{11}\vert^2$. 

\begin{widetext}

\begin{figure}
\includegraphics*[width=2.0in,keepaspectratio]{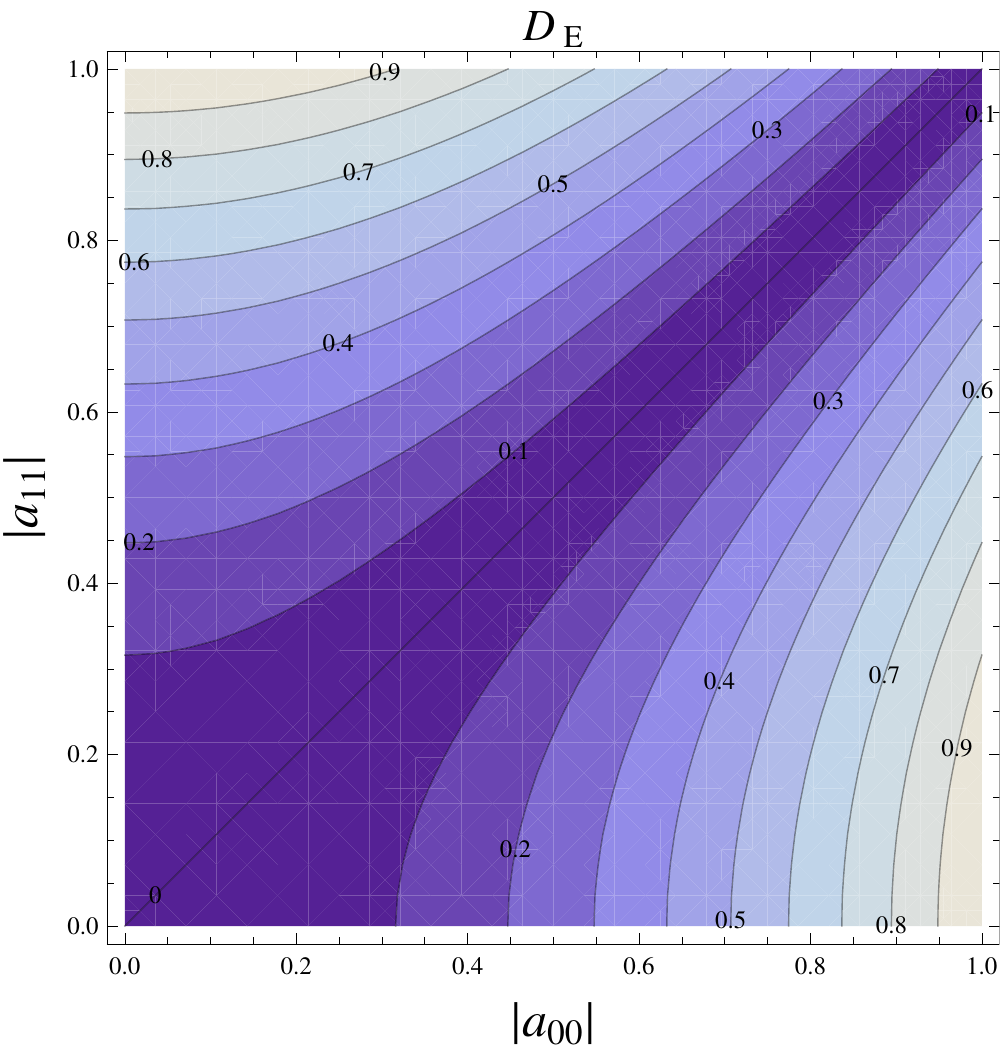}  \includegraphics*[width=2.0in,keepaspectratio]{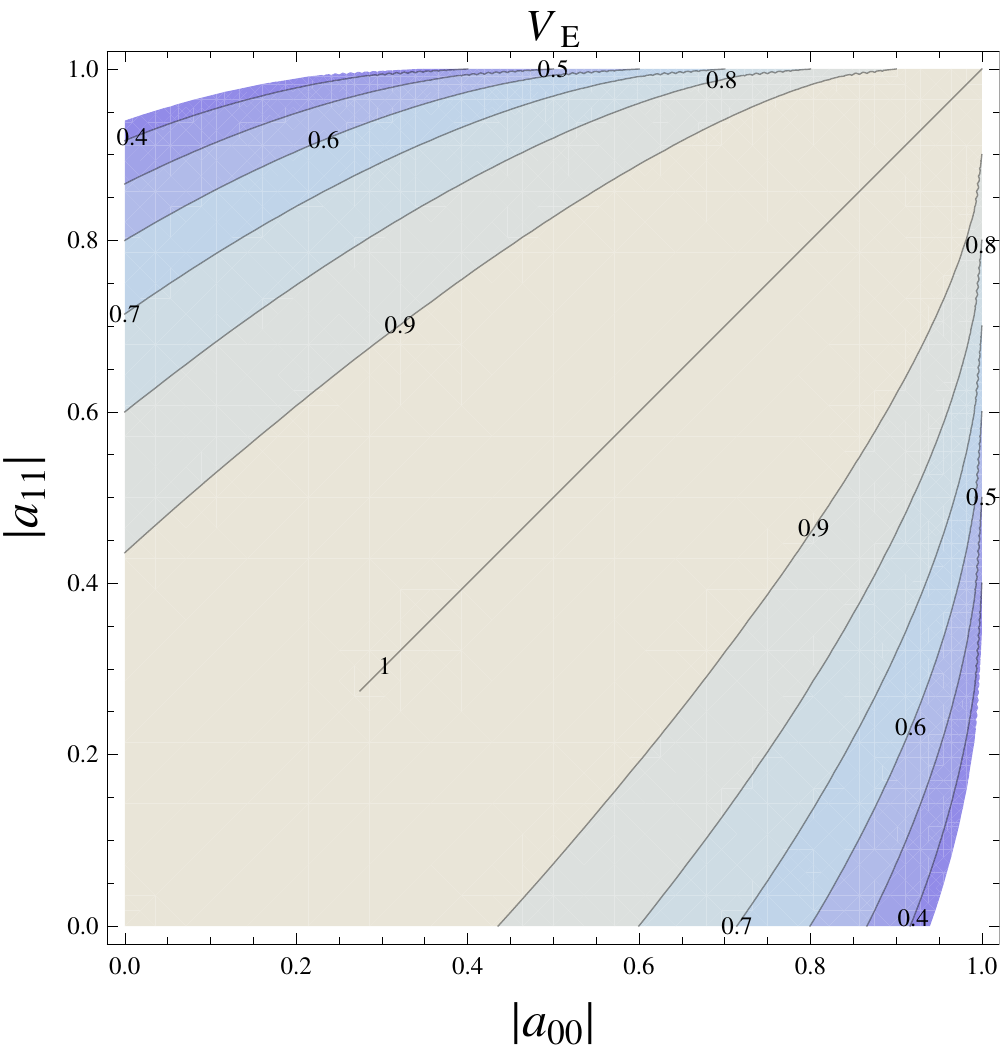}
\includegraphics*[width=2.0in,keepaspectratio]{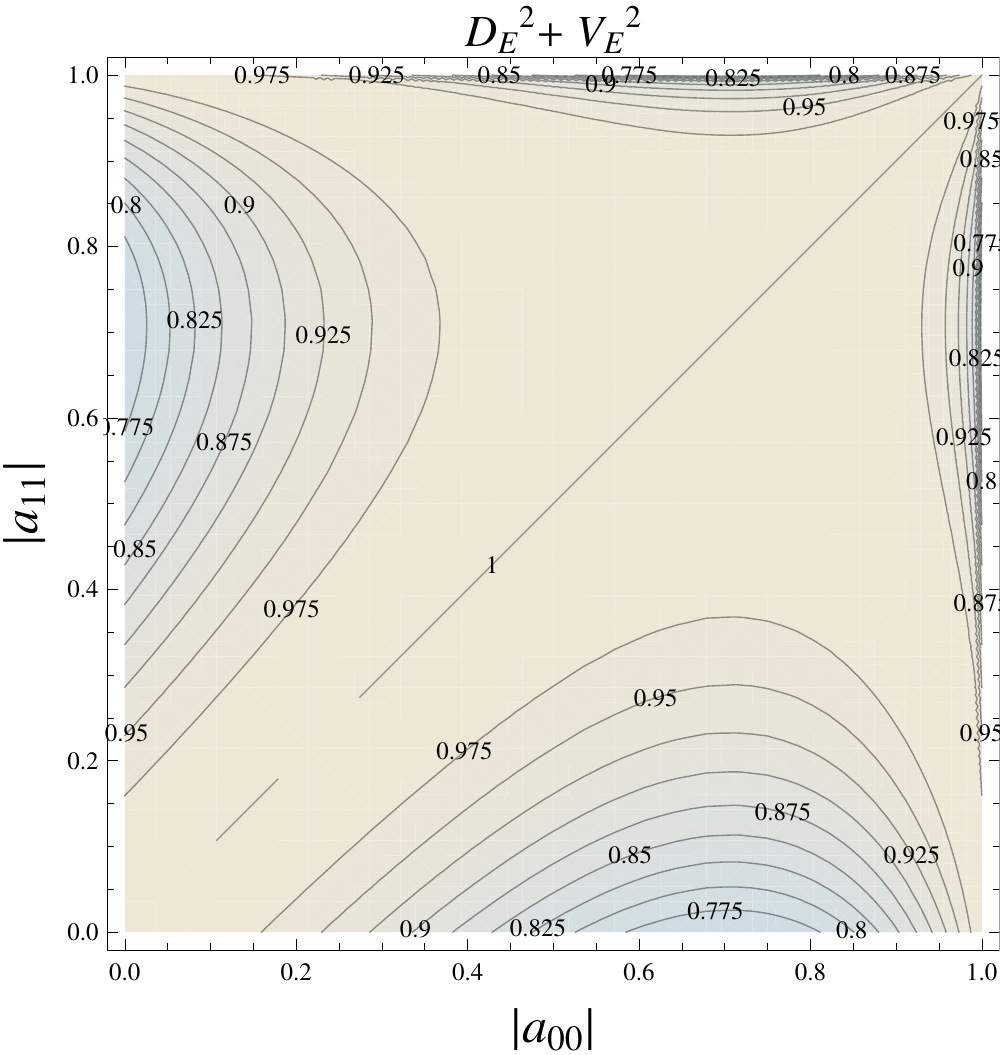}
\caption{Contour plots of {\em extended distinguishability} ${\cal D}_E$, {\em extended visibility} ${\cal V}_E$ (see (\ref{1de}) and (\ref{1ve})) and  ${\cal D}^2_E+{\cal V}^2_E$,  as  functions of $|a_{00}|,\ |a_{11}|$, the parameters of the initial spin preparation. It is clearly seen that the which-way distinguishability  ${\cal D}_E$ and fringe visibility ${\cal V}_E$ obey the duality relation ${\cal D}^2_E+ {\cal V}^2_E\leq 1$.}
\end{figure}
\end{widetext}
The extended visibility ${\cal V}_E$ (see (\ref{fv})) gets simplified to the following:  
\begin{equation} 
\label{1ve}
{\cal V}_E=\vert a_{00}\vert\,  \vert a_{11}\vert\, +\, \sqrt{1- \vert a_{00}\vert^2}\, \sqrt{1- \vert a_{11}\vert^2}.  
\end{equation}

In Fig.~1 we have plotted the {\em extended which-way distinguishability} ${\cal D}_E$, the {\em extended fringe visibility} ${\cal V}_E$ and ${\cal D}^2_E+{\cal V}^2_E$ as a function of the initial spin state parameters  $\vert a_{00}\vert,\ \vert a_{11}\vert$. It is clearly seen that the extended distinguishability, visibility are complementary to each other and they obey the trade-off relation ${\cal D}^2_E+{\cal V}^2_E\leq 1$.

In the absence of the ancilla $D'$, we find the detector states,  
$\rho^{(0)}_{D}={\rm Tr}_{D'}\, [\rho^{(0)}_{DD'}]=\frac{1}{2}\left(\begin{array}{ll} 1 & 0 \\ 0 & 1 \end{array}\right),$ 
$\rho^{(1)}_{D}={\rm Tr}_{D'}\, [\rho^{(0)}_{DD'}]=\frac{1}{2}\left(\begin{array}{ll} 1 & 0 \\ 0 & 1 \end{array}\right),$
are perfectly indistinguishable leading to which-way distinguishability  ${\cal D}=0$ when ancilla is not taken into consideration.

We also evaluate the generalized fringe visibility introduced in Ref.~\cite{Ban} (see (\ref{vg})) in this example. From the final state density operator 
$\rho^{\rm (fin)}_{QS}$ (see (\ref{finqs})) of the quanton path-spin system, we identify that the action of the quantum channel $\Lambda_{01}$ on spin states is to kill the diagonal elements of the spin operator. More explicitly,       
\begin{equation}
\Lambda_{01}(\vert \alpha\rangle_S\langle \alpha '\vert)=\left\{ \begin{array}{l} 0,\  {\rm if}\ \alpha=\alpha ' \\ 
  1,\  {\rm if}\ \alpha \neq \alpha ' \end{array}\right. ,   
\end{equation} 
where $\alpha, \alpha '=0,1$. We simplify (\ref{vg}) to obtain the generalized fringe visibility
\begin{equation}
{\cal V}_G=\vert a_{00}\vert\,  \vert a_{11}\vert\, +\, \sqrt{1- \vert a_{00}\vert^2}\, \sqrt{1- \vert a_{11}\vert^2},
\end{equation}   
which matches exactly with the extended visibility ${\cal V}_E$ of (\ref{1ve}). 

It may be noted that even though the which-way distinguishability ${\cal D}$ (obtained when the ancilla degree of freedom is ignored) is zero, the generalized visibility ${\cal V}_G$ does not take its maximum value 1 in this example. Variation of ${\cal V}_G$  does indeed reveal a leakage of which-way information. However, the  trade-off relation turns out to be that between the visibility and the which-way information captured by  the extended distinguishability ${\cal D}_E$ -- and not the one assimilated through ${\cal D}$.      

Next, we consider the initial  quanton path-spin to be prepared in the state  $\vert\zeta_{QS}\rangle^{\rm (in)}= 
 \frac{1}{2}[\vert 0\rangle_Q +\vert 1\rangle_Q]\otimes \vert 0\rangle_S$ and  the detector-ancilla state is initially prepared in the maximally entangled state $\vert \Psi\rangle_{DD'}$ given by (\ref{ee'}). The unitary interaction between the quanton spin and detector is chosen to be,  
\begin{eqnarray}
U^{(0)}_{SD}= \left( \begin{array}{llll}
0 & 0 & 0 & 1 \\ 
0 & 0  & 1 & 0 \\
0 & 1  & 0 & 0 \\
1 & 0 & 0 & 0 \end{array}
\right)\ \ 
U^{(1)}_{SD}= \left( \begin{array}{llll}
0 & 0 & 0 & 1 \\ 
0 & 0  & e^{-i\phi} & 0 \\
0 & e^{i\phi}  & 0 & 0 \\
1 & 0 & 0 & 0 \end{array}
\right)
\end{eqnarray} 
which are given in the spin-detector basis states $\{\vert 0\rangle_S\, \vert 0\rangle_D,\, \vert 0\rangle_S\, \vert 1\rangle_D,$  $\vert 1\rangle_S\, \vert 0\rangle_D,$  $\vert 1\rangle_S\, \vert 1\rangle_D\})$.
After the interaction, the detector-ancilla states associated with the paths $0$, $1$ of the quanton are given by, 
\begin{equation}
\rho^{(0)}_{DD'}=\frac{1}{2}\left(\begin{array}{llll}  
0 & 0 & 0 & 0 \\ 
0 & 1 & 1 & 0 \\ 
0 & 1 & 1 & 0 \\ 
0 & 0 & 0 & 0 
 \end{array}\right),\,   \rho^{(1)}_{DD'}=\frac{1}{2}\left(\begin{array}{llll}  
0 & 0 & 0 & 0 \\ 
0 & 1 & e^{-i\phi} & 0 \\ 
0 & e^{i\phi} & 1 & 0 \\ 
0 & 0 & 0 & 0 
 \end{array}\right)
\end{equation}  
The  which-way information extracted via  the extended distinguishability is $ {\cal D}_E=\vert \sin(\phi/2)\vert$,  while the extended fringe visibility is identified to be exactly complementary~\cite{note3} i.e.,   ${\cal V}_E=\vert \cos(\phi/2)\vert$.  It is easy to see that the detector states $\rho^{(i)}_{D}={\rm Tr}_{D'}\,[\rho^{(i)}_{DD'}]=I_S/2,\ \ i=0,1$ are indistinguishable.  

In order to evaluate the generalized visibility ${\cal V}_G$, we  first identify the action of the spin channel $\Lambda_{01}$: 
\begin{widetext}
\begin{eqnarray*}    
\Lambda_{01}(\vert 0\rangle_S\langle 0\vert)&=&\left(\frac{1+e^{i\phi}}{2}\right)\, \vert 1\rangle_S\langle 1\vert,\ \
\Lambda_{01}(\vert 1\rangle_S\langle 1\vert)=\left(\frac{1+e^{-i\phi}}{2}\right)\, \vert 0\rangle_S\langle 0\vert, \\ 
\Lambda_{01}(\vert 0\rangle_S\langle 1\vert)&=&\left(\frac{1+e^{-i\phi}}{2}\right)\, \vert 1\rangle_S\langle 0\vert, \ \ 
\Lambda_{01}(\vert 1\rangle_S\langle 0\vert)=\left(\frac{1+e^{i\phi}}{2}\right)\, \vert 0\rangle_S\langle 1\vert. 
\end{eqnarray*} 
\end{widetext}
We thus find that the generalized visibility (\ref{fv}) reduces to ${\cal V}_G=\vert \cos(\phi/2)\vert$, which is equal to the extended visibility ${\cal V}_E$ in this example too. 

Even here, the variation of ${\cal V}_G$ would indicate leak-out of which-path information -- but it is not retrievable from the detector alone -- thus bringing out the significance of detection using the extended detector-ancilla system. 

In the third example, we consider the unitary interaction between the quanton and the detector  to be 
\begin{eqnarray}
U_{QSD}&=&\sum_{i=0,1}\, \vert i\rangle_Q\langle i\vert \otimes U^{(i)}_{SD},\ 
U^{(i)}_{SD}=e^{-i\frac{\theta_i}{2}\, \sigma_z\otimes \sigma_x}  
\end{eqnarray}   
where $\sigma_z\otimes \sigma_x =\left(\vert 0\rangle_S\langle 0\vert -\vert 1\rangle_S\langle 1\vert\right)\otimes \left(\vert 0\rangle_D\langle 1\vert +
\vert 1\rangle_D\langle 0\vert\right)$. 

With an initial quanton path-spin state $\vert\, \zeta\rangle_{QS}=\left(\frac{\vert 0\rangle_Q+\vert 1\rangle_Q}{\sqrt{2}}\right)\, \left(\frac{\vert 0\rangle_S+\vert 1\rangle_S}{\sqrt{2}}\right)$ and the maximally entangled detector-ancilla state (\ref{ee'}), we find that 
\begin{equation}
\rho^{(i)}_{DD'}=\frac{1}{2}\left(\begin{array}{llll}  
\cos^2\left(\frac{\theta_i}{2}\right) & 0 & 0 & \cos^2\left(\frac{\theta_i}{2}\right) \\ 
0 & \sin^2\left(\frac{\theta_i}{2}\right) & \sin^2\left(\frac{\theta_i}{2}\right) & 0 \\ 
0 & \sin^2\left(\frac{\theta_i}{2}\right) & \sin^2\left(\frac{\theta_i}{2}\right) & 0 \\ 
\cos^2\left(\frac{\theta_i}{2}\right) & 0 & 0 & \cos^2\left(\frac{\theta_i}{2}\right) 
 \end{array}\right)
\end{equation}    
are the detector-ancilla final states, corresponding to the quanton paths $i=0,1$. Discrimination of these detector-ancilla states results in 
${\cal D}_E=\frac{1}{2}\, \vert\vert \rho^{(0)}_{DD'}-\rho^{(1)}_{DD'} \vert\vert=\vert \cos^2\left(\frac{\theta_0}{2}\right)- 
\cos^2\left(\frac{\theta_1}{2}\right)\vert $.  The extended fringe visibility is found to be ${\cal V}_E=\left\vert \cos\left(\frac{\theta_0}{2}\right)\, 
\cos\left(\frac{\theta_1}{2}\right)\right\vert + \left\vert \sin\left(\frac{\theta_0}{2}\right)\, 
\sin\left(\frac{\theta_1}{2}\right)\right\vert$. In this example too, the detector states $\rho^{(i)}_{D}={\rm Tr}_{D'}[\rho^{(i)}_{DD'}]=I_D/2$ after the interaction are indistinguishable and they are incapable of retrieving the which-way information. 

We find that the generalized visibility reduces to the extended visibility in this example also. In order to see this, we first identify  the operation of the spin interaction channel $\Lambda_{01}$: 
\begin{eqnarray*}    
\Lambda_{01}(\vert 0\rangle_S\langle 0\vert)&=&\cos\left(\frac{\theta_0-\theta_1}{2}\right)\, \vert 0\rangle_S\langle 0\vert, \\
\Lambda_{01}(\vert 0\rangle_S\langle 1\vert)&=&\cos\left(\frac{\theta_0+\theta_1}{2}\right)\, \vert 0\rangle_S\langle 1\vert, \\
\Lambda_{01}(\vert 1\rangle_S\langle 0\vert)&=&\cos\left(\frac{\theta_0+\theta_1}{2}\right)\, \vert 1\rangle_S\langle 0\vert, \\ 
\Lambda_{01}(\vert 1\rangle_S\langle 1\vert)&=&\cos\left(\frac{\theta_0-\theta_1}{2}\right)\, \vert 1\rangle_S\langle 1\vert.  
\end{eqnarray*} 
After simplification of (\ref{vg}), we obtain the generalized visibility ${\cal V}_G=\left\vert \cos\left(\frac{\theta_0}{2}\right)\, 
\cos\left(\frac{\theta_1}{2}\right)\right\vert + \left\vert \sin\left(\frac{\theta_0}{2}\right)\, \sin\left(\frac{\theta_1}{2}\right)\right\vert$, which agrees perfectly  with the extended visibililty.     

It is pertinent to point out here that in both the first and the third examples the trade-off between the which way information and the visibilities turn out to be identical (which is evident by the parametrization $\vert a_{00}\vert =\left\vert \cos\left(\frac{\theta_0}{2}\right)\right\vert$ and $\vert a_{11}\vert =\left\vert \cos\left(\frac{\theta_1}{2}\right)\right\vert$). While in the first example, the trade-off is realized for different intial spin preparations (by varying the initial spin parameters $\vert a_{00}\vert, \ \vert a_{11}\vert$), analogous trade-off is brought out by varying the parameters of the interaction channel.

\section {Conclusions}

 In a recent work on the interference of a quanton in a two path interferometer, Banaszek et. al.~\cite{Ban} showed that  a control over  the leakage of  which-way information can be achieved by appropriate initial preparation of  the internal spin state of the quanton.  In this work, we extended this analysis and showed  that with the help of an entangled detector-ancilla system,  the amount of which-way information could get enhanced beyond that discerned solely from the detector. Our analysis  generalizes the  trade-off relation between  the which-way distinguishability  and the fringe visibility,   when the detector is equipped with an ancillary degree of freedom. We considered three different  examples of  interaction between the quanton and the detector to demonstrate that the extended which-way distinguishability ${\cal D}_E$  can assume non-zero values even when the  distinguishability  ${\cal D}$ inferred only by the detector  vanishes. In the meanwhile, we also find that the extended visibility ${\cal V}_E$ and the generalized visibilty~\cite{Ban} ${\cal V}_G$  agree identically with each other in these examples. These illustrative examples analyzed here, reveal that there are instances where the detector fails to gain any information on which-way distinguishability, but the corresponding generalized visibility does not attain its maximum value 1. In fact, variation of the generalized visibility $0\leq {\cal V}_G\leq 1$ --  even when  the which-way information transferred to the detector is completely erased  --  draws striking attention. These examples indeed bring forth the need for our extended analysis to explore how leakage of which-way information is captured by the entangled detector-ancilla system, but not by the detector state alone. However, the agreement of the extended  visibility with generalized visibilty in these specific examples appears to be conincidental. It would be of interest to investigate how both these fringe visibilities  are related to each other in general. Furthermore, we leave open the question,   {\em "is it possible to represent the extended fringe visibility ${\cal V}_E$ in terms of quantities that could be controlled at the stage of preparation of quanton internal spin state, so that one can prevent the which-way information leak-out to the combined detector-ancilla system?"}, for further exploration.  
                                                                                                                                                                                                                                                                                                                                                            
                                                                                                                      \begin{center}                                                                                                                    
																																																									{\bf ACKNOWLEDGEMENTS} 
\end{center} 																																																																																																																					We thank Konrad Banaszek for insightful discussions. J.~P acknowledges support from UGC-BSR, Government of India.


\begin{thebibliography}{0}
\bibitem{note} The  channels $\Phi_0$, $\Phi_1$ are identified as follows:  $\Phi_i(\rho^{\rm (in)}_D)= 2\, _Q\langle\, i\vert {\rm Tr}_{S}\, [U_{QSD}\, 
\rho^{\rm (in)}_{QS}\otimes \rho^{\rm (in)}_D\,  U^\dag_{QSD}]\vert i\rangle_Q$ where the unitary interaction is of the form  
$U_{QSD}=\vert 0\rangle _Q\langle 0\vert\, \otimes U^{(0)}_{SD}+\vert 1\rangle _Q\langle 1\vert\, \otimes U^{(1)}_{SD}$ such that quanton path does not get switched between interferometer arms due to the interaction. $U^{(0)}_{SD}$, $U^{(1)}_{SD}$ are unitary operators acting on the spin and detector systems.
\bibitem{note2} It is generally known~\cite{kit} that optimal distinguishability between a pair of channels does not require an ancillary state that is larger 
than the input space of the channels.
\bibitem{note3} The detector-ancilla states $\rho^{(i)}_{DD'},\ i=0,\, 1$ are pure and thus equality sign holds in the relations between trace distance and fidelity (see (\ref{df})).
\end{thebibliography}
\end{document}